# MOCVD synthesis of compositionally tuned topological insulator nanowires

**Loren D. Alegria[1], Nan Yao[2], and Jason R. Petta[*,1]**

[1] Department of Physics, Princeton University, Princeton, NJ 08544, USA
[2] Princeton Institute for the Science and Technology of Materials, Princeton University, Princeton, NJ, USA

Device applications involving topological insulators (TIs) will require the development of scalable methods for fabricating TI samples with sub-micron dimensions, high quality surfaces, and controlled compositions. Here we use Bi-, Se-, and Te-bearing metalorganic precursors to synthesize TIs in the form of nanowires. Single crystal nanowires can be grown with compositions ranging from $Bi_2Se_3$ to $Bi_2Te_3$, including the ternary compound $Bi_2Te_2Se$. These high quality nanostructured TI compounds are suitable platforms for on-going searches for Majorana Fermions [1, 2].

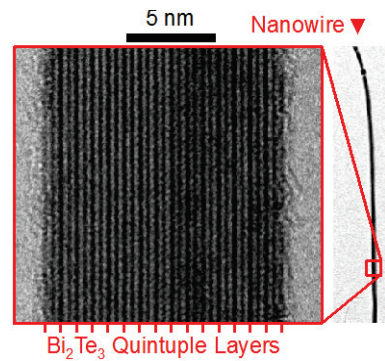

High resolution transmission electron microscope image of a $Bi_2Te_3$ nanoribbon.

**1 Introduction** In conventional insulators, the conduction and valence bands are separated by a band gap, a region in energy space that does not support energy bands. Topological insulators (TIs) are a recently discovered state of matter that are characterized by unique electronic and spin states [3]. Like conventional insulators, TIs are insulating in the bulk. However, unlike conventional insulators, TIs support conducting surface states that traverse the bulk energy gap [4, 5]. These surface states are robust against backscattering due to the large spin-orbit coupling inherent to TI compounds, which locks the spin of the electron to its momentum. For these reasons, there is widespread interest in developing low dissipation electronic devices based on TIs [3, 6-9].

$Bi_2Se_3$ and $Bi_2Te_3$ have long been known as distinctively layered semiconductors with narrow band gaps and exceptional thermoelectric performance [10-12]. Their recent identification as TIs with topologically protected, spin-polarized surface bands has earned them a place among other layered materials being researched for next-generation device architectures, such as graphene nanoribbons [3, 5, 13]. Transport through topological surface states is particularly attractive for possible applications involving spin transport in solids, but can be degraded by surface disorder and parallel conduction through bulk channels. This motivates the study of high quality TI nanostructures with large surface-to-volume ratios [6, 14, 15]. Known methods for growing TI nanowires, mostly developed within the pursuit of improved thermoelectric elements, often produce thick or polycrystalline nanowires with rough surfaces [16]. In contrast, high performance applications of semiconductor nanowires, such as photovoltaics and single electron quantum devices, rely heavily on metalorganic chemical vapor deposition (MOCVD) for the precise growth of nanowires over large areas [1, 17].

**2 Sample Growth and Characterization** In this work, we demonstrate the MOCVD growth of highly ordered TI nanowires, including the advanced ternary compound $Bi_2Te_2Se$ [18]. With this synthesis approach, we grow crystalline $Bi_2Te_3$ nanowires with diameters as small as 12 nm, which have long been sought for thermoelectric research [10, 19]. We further demonstrate the nanowire growth of $Bi_2Se_3$ and ternary compositions which provide a route to chemical depletion of bulk carriers [20]. In transmission electron microscopy (TEM) studies, Bi$_2$Te$_3$ nanowires display the highest degree of surface order. Low temperature magnetotransport measurements on Bi$_2$Te$_3$ nanowires and platelets indicate that these TI nanostructures support long electronic phase coherence lengths and have strong spin-orbit coupling.

* Corresponding author: e-mail petta@princeton.edu, Phone: +1 609 258 1173

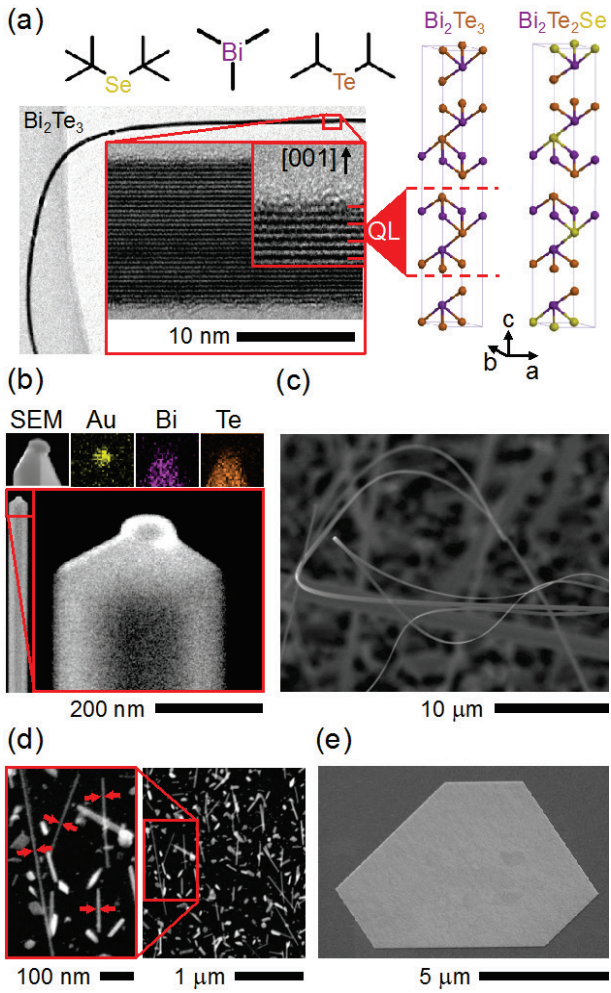

**Figure 1** MOCVD growth of Bi$_2$Te$_3$ nanostructures. (a) The precursors DTBSe, TMBi, and DiPTe. (right) The unit cells of Bi$_2$Te$_3$ and Bi$_2$Te$_2$Se span three quintuple layers (QL) which are directly visible in nanowires imaged via TEM along the a-b plane. (b) Au catalysts are found at the ends of nanowires, but in contrast with standard VLS growth, the Bi$_2$Te$_3$ nanowire diameters are typically larger than the catalyst. (c) Long nanowires, prepared by increasing the growth time to 1 hr. (d) 5 nm Au seed particles, in combination with low growth temperatures (here, 260 °C) allow for ultra-thin nanowires, as shown in this SEM image of the substrate following growth. In the left image, the pairs of red arrows are separated 12 nm. (e) Growth on SiO$_2$ without catalysts produces smooth, flat platelets with thickness < 20 nm.

Bi$_2$Se$_3$ and Bi$_2$Te$_3$ have a complex crystal structure consisting of van der Waals bonded quintuple layers, as depicted in Fig. 1. The development of these compounds as high quality TIs has focused on reducing the level of bulk conduction, which is believed to be due to the formation of vacancies and anti-site defects [3]. The degree to which defects form during growth from the vapor phase depends on the reactivity of the vapor sources: large atom clusters in the vapor are less mobile and less able to integrate into the growing crystal structure [21]. In contrast to the molecular beam epitaxy techniques employed widely in the field of TI growth, MOCVD produces an atomic vapor since each precursor molecule bears a single metal atom, leading to a highly reactive vapor [22]. We therefore pursue the growth of Bi$_2$Te$_3$ and Bi$_2$Se$_3$ using the precursors trimethyl bismuth (TMBi), di-tert-butyl selenium (DTBSe), and diisopropyl tellurium (DiPTe) [Fig. 1(a)] which have been previously developed for the deposition of GaAs$_{1-x}$Bi$_x$, ZnSe, and CdTe [23, 24] and used separately for the growth of Bi$_2$Te$_3$ and Bi$_2$Se$_3$ films [11, 25]. The primary growth variables are the substrate growth temperature, $T_g$, the group VI/V precursor partial pressure ratios, $r_{Se} = p_{DTBSe}/p_{TMBi}$ and $r_{Te} = p_{DiPTe}/p_{TMBi}$, and the gold catalyst particle diameter (60 nm unless otherwise noted).

The growth substrate consists of an RCA-cleaned Si (100) chip prepared with Au nanoparticles with a density of about 1 μm$^{-2}$. A 100 Torr, 600 sccm flow of research grade H$_2$ carrier gas is established in the growth chamber, and the substrate is cleaned by heating to 600 °C for 60 s. The temperature is then reduced to 420 °C, at which point dilute vapors of the group VI precursors are introduced into the H$_2$ carrier gas stream, producing partial pressures $p_{DTBSe}$ and $p_{DiPTe}$ in the growth chamber. After 120 s, the temperature is set to the growth temperature $T_g$ and the Bi precursor ($p_{TMBi} = 7 \times 10^{-6}$ atm) is introduced. We find that the three precursors integrate without prereaction and allow for the growth of nanowires of Bi$_2$Se$_3$ and Bi$_2$Te$_3$ within specific parameter ranges, and nanowires of the ternary compounds at the intersection of these ranges. After 1800 s, the growth is terminated by halting the TMBi flow and allowing the substrate to cool to 125 °C in continued group VI flow. Typical growth parameters are plotted as a function of time in Fig. 2(a), and departures from these standard conditions are noted in the text.

DiPTe is used for the deposition of tellurium. The low decomposition temperature of this precursor allows a wide range of Bi$_2$Te$_3$ growth conditions, producing a variety of Bi$_2$Te$_3$ products, as summarized in Fig. 1 [26]. We find that nanowire growth dominates when catalysts are present on the substrate either in the form of colloidal gold particles or gold films [Figs. 1(a-d) show such nanowires]. A 15 nm thick nanowire grown this way is studied via TEM in Fig. 1(a). The nanowire is twisted on the TEM grid, revealing an approximately square cross section. In the high-resolution image, the wire can be seen to consist of a continuous belt of 15 stacked quintuple layers. The [001] interface to vacuum at the top and bottom of this image shows no evidence of oxidation despite atmospheric exposure.

Larger colloidal Au particles result in larger diameter nanowires, but we find it necessary to increase the temperature to obtain significant yields of large nanowires. Figure 1(b) shows the endpoint of a thick nanowire typical of growth at $r_{Te}$ = 4 and $T_g$ = 350 ºC and catalyzed with a 60 nm Au particle. The narrowest structures are achieved with $r_{Te}$ = 4 and $T_g$ = 260 ºC using 5 nm seed particles, producing nanowires with an average thickness of ~12 nm, as shown in Fig. 1(d).

The growth mechanism of $Bi_2Te_3$ nanowires is investigated by energy dispersive spectroscopy (EDS) compositional mapping of the nanowire endpoints, as illustrated in Fig. 1(b). The terminal seed particles consist predominantly of Au and the diameters of the nanowires typically exceed the catalyst particle diameters, so it is probable that some non-catalyzed sidewall growth occurs. In very long nanowires, similar to the one shown in Fig. 1(c), terminal particles are often absent, or replaced by a disoriented $Bi_2Te_3$ crystallite (in this panel, the growth duration was increased to one hour, with $r_{Te}$ = 4, and $T_g$ = 350 ºC to obtain long wires). Flat platelets (with thickness < 20 nm) grow in the absence of catalysts, as shown in Fig. 1(e).

To explain the coexistence of catalyst-free and catalyzed nanowire growth, we note that early studies of the vapor-liquid-solid growth mechanism demonstrated GaAs nanowire growth from non-stoichiometric regions of GaAs crystals, without the presence of an added catalyst particle [27]. $Bi_2Se_3$-type compounds may similarly grow without an external catalyst from a Bi-rich region. On the other hand, in Au-catalyzed GaAs nanowire growth, the bulk phase diagram is predictive of the Au-Ga melt [28]. In contrast to Au-Ga, Au-Bi does not form a eutectic until the melt is almost entirely Bi [29]. We would therefore hypothesize that for catalyzed $Bi_2Te_3$ growth, Au remains a solid, nearly pure particle, which is consistent with our observation of terminal Au particles with low Bi content, although in-situ studies would be definitive [30].

$Bi_2Se_3$ nanowires are grown using the precursor DTBSe. In contrast to growth using diethyl selenium (DESe), nanowire growth from DTBSe can occur at much reduced temperatures, widening the range of suitable growth conditions and thereby allowing for nanowire growth of ternary compounds [24, 31]. We vary the temperature and precursor ratios in a series of growth runs, and determine the resulting elemental compositions using EDS. With $r_{Se}$ = 20 fixed, growth at $T_g$ = 330 ºC produces pure Bi spheres and at $T_g$ = 350 ºC, 360 ºC, and 370 ºC produces crystalline nanowires of the Bi-rich phases $Bi_{66\%}Se_{34\%}$, $Bi_{50\%}Se_{50\%}$, and $Bi_{51\%}Se_{49\%}$ respectively (±1 at.%). We conclude from this series that the decomposition of the DTBSe occurs near 350 ºC and that sufficiently high $r_{Se}$ will likely produce $Bi_2Se_3$ nanowires at temperatures 350 - 370 ºC. Indeed, when $r_{Se}$ is then increased at fixed $T_g$ = 360 ºC, $Bi_{46\%}Se_{54\%}$ nanowires grow at $r_{Se}$ = 35, and nanowires of the stoichiometric composition $Bi_2Se_3$ ($Bi_{40\%}Se_{60\%}$) grow at $r_{Se}$ = 70. The EDS spectrum for the $Bi_2Se_3$ nanowires is shown in Fig. 2.

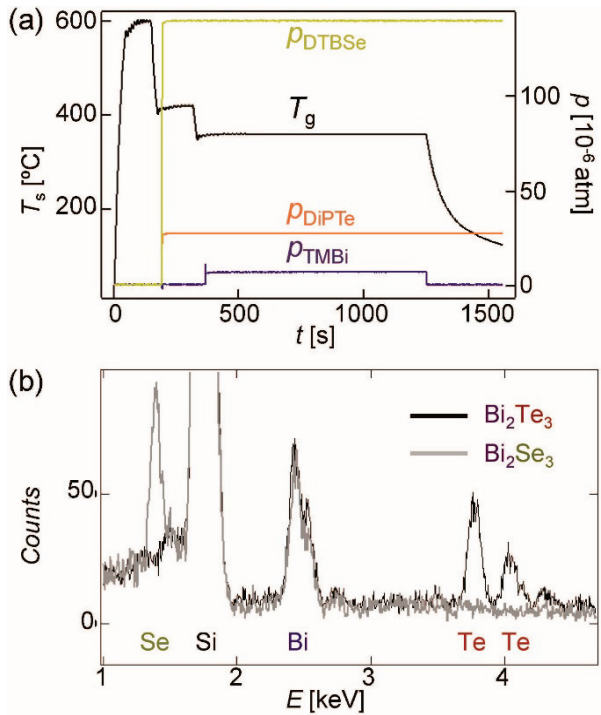

**Figure 2** Growth procedure for $Bi_2Te_3$ and $Bi_2Se_3$ nanowires. (a) The substrate temperature ($T_s$) and precursor partial pressures ($p$) are plotted as a function of time. The use of the precursor DTBSe enables $Bi_2Se_3$ growth at low temperatures that are compatible with $Bi_2Te_3$ growth using DiPTe. (b) EDS spectra of $Bi_2Se_3$ and $Bi_2Te_3$ nanowires grown by MOCVD on Si substrates (30 s of data collection). In both cases, we determine a 40%/60% composition (±1 at.%).

Overlap in the suitable conditions for $Bi_2Te_3$ and $Bi_2Se_3$ nanowire growth opens the possibility of growing ternary nanowires. Control of the defects in $Bi_2Te_2Se$ makes it possible to grow highly insulating crystals of this composition [18, 32]. To achieve the growth of ternary nanowires, we set $T_g$ = 360 ºC and vary the precursor pressures. In practice, we fix $p_{TMBi}$ = 5 × 10$^{-6}$ atm, $r_{Te}$ = 4, and vary the Se richness from $r_{Se}$ = 20 to 28. The nanowire products are single crystalline and similar to the $Bi_2Te_3$ nanowires, as shown in the TEM images of Fig. 3. The [Bi]/([Se] + [Te]) ratio in the nanowires is consistently 2/3, but $r_{Se}$ controls the atomic ratio of [Se]/[Te] in the nanowires. As shown in the inset of Fig. 3(a), a roughly linear dependence of [Se]/[Te] = 0.4 - 0.8 is observed as a function of $r_{Se}/r_{Te}$ = 5 - 7. The composition sought for transport experiments ($Bi_2Te_2Se$) is obtained with $r_{Se}/r_{Te}$ = 5.8 [Fig. 3(a)]. The [001] surfaces of such nanowires are studied using TEM, as shown in Fig. 3(b).

**3 Transport Measurements** In addition to TEM studies, transport measurements indicate that the nanowires are of high quality. Specifically, we measure the coherence and spin-orbit lengths via magnetoconductance measurements. The results indicate low defect densities, strong spin-orbit coupling, and long coherence lengths in $Bi_2Te_3$, all of which are encouraging for applications utilizing coherent transport of carriers or current-driven spin polarization. Devices are fabricated by transferring nanowires and platelets to an oxidized (100 nm of $SiO_2$) p++ Si substrate. Nanostructures are individually contacted using electron beam lithography and liftoff. To achieve low resistance contacts, a low energy Ar ion etch is performed to the contact region prior to thermal evaporation of 5 nm Ti and 60 nm Au.

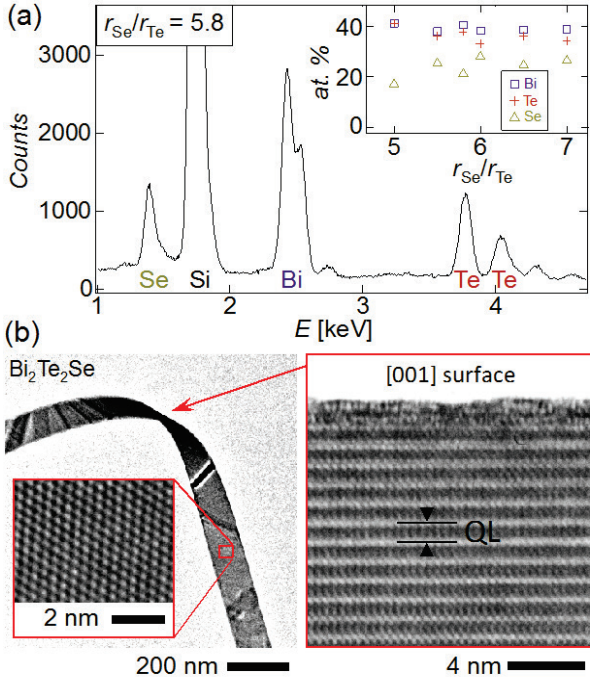

**Figure 3** Growth of ternary nanowires via MOCVD. (a) The EDS spectrum of a sample grown under conditions $r_{Se}/r_{Te}$ = 5.8, which indicates a $Bi_{41\%}Te_{38\%}Se_{21\%}$ composition (±1 at.%) (120 s of data collection). In the inset, the product composition is plotted as a function of $r_{Se}/r_{Te}$. (b) TEM imaging of $Bi_2Te_2Se$ nanowires along the c-axis (left) and a-b plane (right) reveals the single crystal nature of the nanowires. The surface of the sample remains highly ordered. The quintuple layer spacing indicated for $Bi_2Te_2Se$ is 1.00 nm.

Angle-resolved photoemission spectroscopy experiments on TIs show that strong spin-orbit coupling locks the electron spin to its momentum [4, 5]. We probe the strength of the spin-orbit coupling in our TI nanostructures by performing magnetotransport measurements. The conductance, g, is plotted as a function of magnetic field, B, in Fig. 4 for n-type $Bi_2Te_3$ samples (grown at $r_{Te}$ = 4, $T_g$ = 280 ºC). The sharp peak at low field is due to weak anti-localization (WAL). The reproducible structure at higher fields is due to universal conductance fluctuations (UCF) [33].

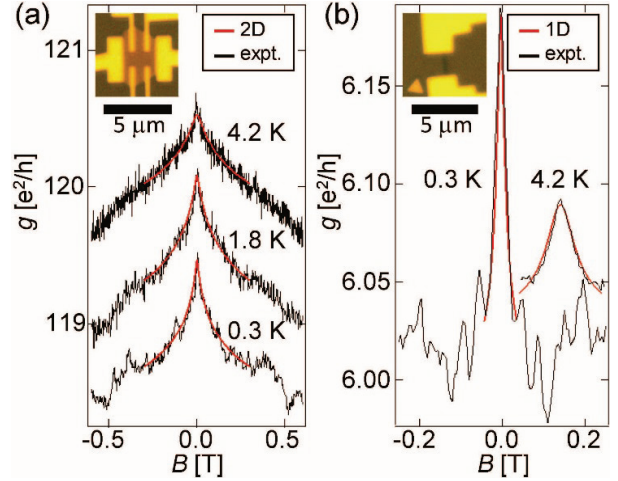

**Figure 4** WAL in nanoscale $Bi_2Te_3$ samples. The conductance is plotted as a function of perpendicular magnetic field. (a) The four-probe conductance in a thin platelet of $Bi_2Te_3$ is fitted to the 2D model of WAL (red) as described in the text. The resulting $l_\varphi$ are 270 nm, 360 nm, and 480 nm at 4.2 K, 1.8 K, and 0.3 K. (b) The conductance in a $Bi_2Te_3$ nanowire device is fit to a 1D model of WAL (red) with resulting $l_\varphi$ of 220 nm and 670 nm at 4.2 and 0.3 K. The 4.2 K data are offset from zero field for clarity.

WAL occurs in materials with large spin-orbit coupling. The strength of the spin-orbit coupling can be characterized in terms of the spin-orbit length, $l_{so}$. In two-dimensional electron gas systems formed in GaAs/AlGaAs heterostructures, $l_{so} \sim 8$ μm [34]. For comparison, the much heavier semiconductor compound InAs has $l_{so} \sim 100$ nm [35]. In still stronger spin-orbit materials, such as $Bi_2Te_3$, the spin-orbit length is much less than the phase coherence length $l_\varphi$. For $l_{so} < l_\varphi$, a small magnetic field suppresses the set of anti-localized, self-intersecting orbits, causing a conductance peak at low field [36, 37]. The functional form of this peak, $\Delta g(B)$, depends on the geometry of the carrier channels. The vast majority of studies of WAL in TI materials have measured thin film samples, where the width of the channel is much larger than the coherence length ($W > l_\varphi > l_{so}$). We first show that WAL data from a 2D sample grown by MOCVD is consistent with previous reports and then go on to examine 1D samples grown using MOCVD.

For the 2D case, the correction to the conductance depends on magnetic field as

$$\Delta g = \alpha \left(\frac{e^2}{\pi h}\right)\left(\ln\left(\frac{B_\phi}{B}\right) - \Psi\left(\frac{1}{2} + \frac{B_\phi}{2B}\right)\right), \quad (1)$$

where $\Delta g = g(B) - g(B=0)$, $h$ is Planck's constant, $e$ is the elementary charge, α=1/2 for a single 2D channel, and $B_\varphi = h/4el_\varphi^2$ is the phase breaking field [37]. Figure 4(a) shows WAL data from a 2D platelet. We the data using Eq. 1, focusing on the low field regions, where UCF does not strongly influence the conductance and skipping orbits are avoided. Best fits to the data in Fig. 4(a) yield $l_\varphi$ = 270 nm, 360 nm, and 480 nm at 4.2 K, 1.8 K, and 0.3 K.

In contrast, a 1D model must be applied to samples with width much less than $l_\varphi$, as has been clearly demonstrated in GaAs/Al$_x$Ga$_{1-x}$As heterostructures [38]. Here the correction to the conductance is:

$$\Delta g = -\left(\frac{2e^2}{hL}\right)\left[\frac{3}{2}\left(\frac{1}{l_\varphi^2} + \frac{4}{3}\frac{1}{l_{so}^2} + \frac{1}{3}\left(\frac{eWB}{\hbar}\right)^2\right)^{-\frac{1}{2}} - \frac{1}{2}\left(\frac{1}{l_\varphi^2} + \frac{1}{3}\left(\frac{eWB}{\hbar}\right)^2\right)^{-\frac{1}{2}}\right], \quad (2)$$

where $L$ ($W$) is the length (width) of the nanowire [39]. Data from the nanowire shown in Fig. 4(b) (with SEM-determined dimensions $L$ = 3.3 μm, $W$ = 180 nm, and thickness ~ 30 nm) are fit using the 1D theory of Eq. (2). We find best fit values of $l_\varphi$ = 220 nm and 670 nm at 4.2 K and 0.3 K. At both temperatures, we find $l_{so}$ = 40 nm and $W$ = 175 nm which are also taken as free parameters in the fit. The strong signatures of spin-orbit coupling well within the quantum wire regime ($W \ll l_\varphi$) are promising for the generation of Majorana fermions via proximal superconductivity in TI nanowires [2, 7].

**4 Conclusions** In summary, we have demonstrated that a MOCVD growth process can be used to form pristinely ordered nanowires of Bi$_2$Te$_3$, Bi$_2$Se$_3$, and intermediate compounds with long coherence lengths and strong spin-orbit coupling. The exceptional properties of Bi$_2$Te$_3$ nanowires in particular could be useful for devices making use of the strong spin-orbit coupling, such as low-current spin torque transfer devices; for thermoelectric materials making use of surface conductivity added by TI surface states; and for research utilizing both ballistic transport and strong spin-orbit coupling, such as the pursuit of Majorana fermions [1, 7, 8, 10, 14].

**Acknowledgements** We thank Phuan Ong and Bob Cava for valuable discussions. This work was supported by the Packard Foundation, the National Science Foundation through the Princeton Center for Complex Materials (DMR-0819860), the Eric and Wendy Schmidt Transformative Technology Fund, and the Gordon and Betty Moore Foundation.